# Calculation of Optimal Design and Ideal Productivities of Volumetrically-Lightened Photobioreactors using the Constructal Approach

*Jean-François CORNET**

Clermont Université – ENSCCF, EA 3866, Laboratoire de Génie Chimique et Biochimique, BP 10448, F-63000, Clermont-Ferrand, France

Phone : (33) *4-73-40-50-56

Fax : (33) *4-73-40-78-29

J-Francois.CORNET@univ-bpclermont.fr

(*) corresponding author: Laboratoire de Génie Chimique et Biochimique, Bât. Polytech. 24, Avenue des Landais – BP 206 - 63174 AUBIERE Cedex – France

# ABSTRACT

This article examines the optimal design and ideal kinetic performances of volumetrically-lightened photobioreactors (PBR). From knowledge models developed for several years by the author, simple theoretical rules are established at first to define the optimal functioning of solar and artificially-lightened PBR. The constructal approach is then used accordingly, what allows the emergence of the optimal design, or the best lighting structures assembly, in Cartesian and curvilinear geometries, with a privileged treatment for the practical case of the 2D-cylindrical geometry. The obtained results confirm the considerable potential of this approach which is applied here for the first time to the case of the radiant light transfer in participating and reactive media. This enables to define clearly, from a theoretical point of view, the concept of ideal PBR (both for solar or artificial illuminations), which is demonstrated to correspond exactly in most cases to volumetrically-lightened PBR, mainly for the solar DiCoFluV (Dilution Contrôlée du Flux en Volume) concept developed in this article. For this last case, the results of the calculations allow to announce maximal biomass productivities as thermodynamic limits, what can contribute to clarify a today confused debate on this point. The work proposed in this article finally establishes guidelines to conceive more efficient large-scale PBR of any desired geometry and criteria like volume (for artificial illumination) or surface (for solar illumination) maximum productivities and internal or external irradiation.

## 1- Introduction

Photobioreactors (PBR) are specific technologies of reactors with a widespread variety of concepts enabling to cultivate photosynthetic micro-organisms which are today recognized as a possible serious alternative, with other biomass origins, to the exhaustion of the fossil resources. There is thus an important necessity of developing more efficient technologies of PBR and to clarify their respective maximum performances for industrial purposes. This could be done in the framework of rational engineering bases, independently of any economical or profitability considerations, because nobody knows today what will really be the societal role played by photosynthesis and bio-refineries in half a century.

Assuming all other physiological needs controlled at their optimal conditions (pH, temperature, inorganic dissolved carbon…), PBR are processes governed by radiant light transport which determines all the performances and becomes a critical phenomenon for large size PBR conception, explaining the great variety of existing concepts (Carvalho *et al*., 2006; Pulz, 2001). However, this physical problem and its kinetic coupling is often insufficiently understood, leading to numerous misinterpretations in the literature and a conviction that it does not exist rational methods to develop efficient concepts. Contrary to this posture, we discuss in this article a very important concept of varied and highly efficient PBR, namely the volumetrically-lightened PBR (artificially illuminated or solar) which is already existing since the 90s for small scales (An and Kim, 2000; Chen *et al*., 2006; Fleck-Schneider *et al*., 2007, Matsunaga *et al*., 1991; Muller-Feuga *et al*., 1998; Ogbonna *et al*., 1996 and 1999; Ono and Cuello, 2004 and 2006; Suh *et al*., 1998; Suh and Lee, 2003; Takano *et al*., 1992; Zijffers *et al*., 2008), but which was never formalized or optimized so far for large sizes reactors. We define and calculate the optimal design on the strong physical basis of our previous theoretical work (in relation to knowledge models on radiant light transfer and its kinetic and energetic coupling) and using the recent constructal approach develop by Bejan (2000). The problem is described in a general way for three main geometries but with a preference for the practical 2D-cylindrical case, giving the keys to develop many interesting original and efficient concepts in the future.

It is finally demonstrated that for many situations, these optimized concepts correspond to ideal PBR and then give the maximum surface and volume kinetic performances that it is possible to reach by engineering the photosynthesis in PBR. Some specific technological points in connection with these concepts and that it will be necessary to investigate and improve in the future are then discussed at the end of this article.

## 2- Some Theoretical Rules for the Optimal Design of Large Photobioreactors

We demonstrate and discuss here some basic requirements and rules relying on our previous published work in the field of radiant light transfer and PBR engineering, but presented and analyzed in such a way that it becomes easy to understand the good practices (sometimes discussed in the specialized literature but not really formalized on a strong physical basis) in defining appropriate optimal concepts of large size PBR.

We mainly focus our attention on the special case of volumetrically-lightened PBR for which we establish in the following of the article the theoretical superiority of kinetic performances in most of the practical situations.

## 2.1- Physical Limitation by Light Transfer - The $\gamma = 1$ Equation for Optimal Functioning of PBR

It is nowadays widely admitted and recognized that, like any photo-reactive process, PBR performances are governed by the radiant light energy transport inside the culture medium, requiring then to develop specific technologies with high illuminated surface-to-volume ratios (Carvalho *et al*., 2006). Nevertheless, in spite of the great diversity of published or patented concepts and geometries of PBR, the physical phenomenon of light transfer limitation may be generally summarized in a very simple and concise manner, using the concept of working illuminated volume developed by the author a long time ago (Cornet *et al*., 1992). The simple (but representative) case study of a rectangular PBR illuminated by one side is illustrated on Figure 1, enabling to explain that, regarding the light transfer problem in any geometry, only three different situations exist for the reactor functioning. In the first case, the biomass concentration inside the reactor is too low, leading to a true kinetic regime functioning with low performances because the main part of the incident radiant light flux density (LFD) $\overline{q}_0$ is wasted at the rear of the reactor, and additionally, because higher irradiances in the culture medium lead to lower energetic efficiencies for the primarily reactions chains of the photosynthesis (see hereafter). In this case, the working illuminated fraction $\gamma$, defined as the part of the reactor volume having local irradiances higher than the compensation point for photosynthesis $G_C$ (Cornet *et al.*, 1992; Cornet and Dussap, 2009; Takache *et al.*, 2009) - or for the considered rectangular one-dimensional geometry, the extinction length for the irradiance profile - would be hypothetically higher than the actual PBR volume (or length) as labeled on Figure 1 ($\gamma > 1$). At the opposite, in the third case on Figure 1, the biomass concentration is too high, and then a more or less important part of the reactor is in darkness (or with very low irradiances, below the irradiance of compensation $G_C$), corresponding to a value for the working illuminated fraction $\gamma < 1$. This situation also appears as unsatisfactory for the kinetic performances because the dark volume fraction (1- $\gamma$) corresponds at least to a dead zone for the reaction (with however an energetic cost for mixing a non-productive volume) if cultivating cyanobacteria having no short time respiration at obscurity, or to a critical zone in which respiration is predominant for microalgae cultivation with important negative consequences on the mean growth rate of the PBR. Finally, the ideal situation is indeed represented by the second case (see Figure 1) verifying the simple equation $\gamma = 1$ for the working illuminated fraction, i.e. a biomass concentration corresponding exactly to the appearance of the physical limitation by light, the irradiance of compensation being just reached at the rear of the reactor (or in the general case, matching quasi-exactly with the frontiers of the reactor volume). In this condition, all the incident available photons are absorbed in the PBR and used for the photosensitized reactions with an optimal thermodynamic efficiency depending mainly, in a first approximation, on the value of the incident LFD $\bar{q}_0$ alone (Cornet, 2007; Cornet and Dussap, 2009). Thus, it clearly appears that the condition $\gamma = 1$ is a necessary and sufficient condition to ensure maximum kinetic volumetric performances at a given incident LFD, and for any geometry of PBR (we will see later in this article that this condition is not sufficient to ensure maximum surface productivities). This also demonstrates the considerable interest to have the possibility of controlling the biomass concentration in the PBR, i.e. operating in continuous mode, whereas operating in batch always gives very low biomass productivities in any geometry of PBR (the condition $\gamma = 1$ in this case being only fulfilled during some hours of the complete time course).

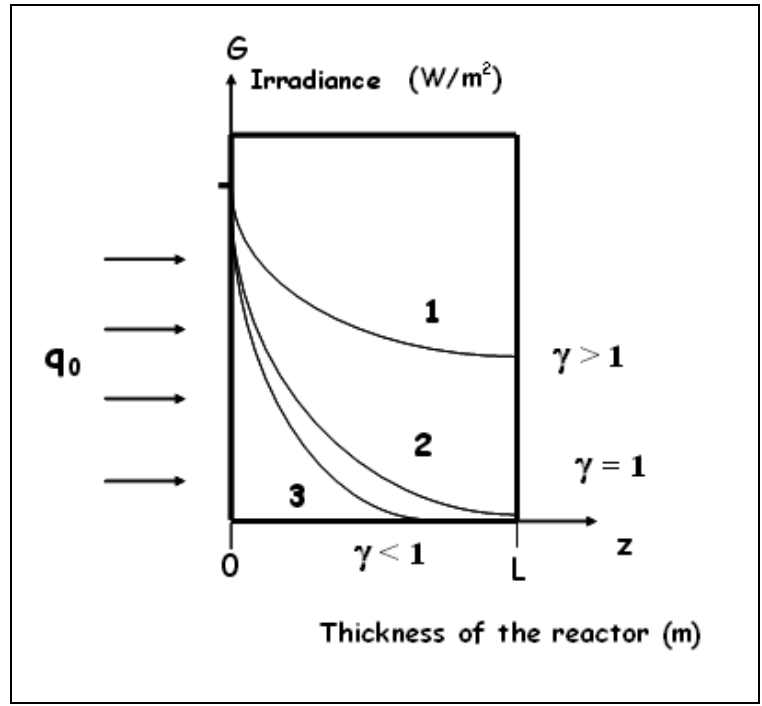

Figure 1: Definition of the working illuminated fraction $\gamma$ in a photobioreactor. The simple case of a plane, one-dimensional rectangular attenuation for irradiance profiles ($G$) is presented with an incident hemispherical light flux density $\bar{q}_0$. The three situations are obtained by respectively increasing the biomass concentration in the reactor. Only the case (2) with $\gamma = 1$ corresponds to an ideal situation for the PBR functioning (see explanations in text).

This so-called working illuminated fraction $\gamma$, as a key but simple parameter, must be easily defined in any geometry and boundary condition for the PBR, from robust radiant light transfer models, requiring to know accurately the radiative properties of the micro-organisms (Berberoglu *et al.*, 2008; Cornet *et al.*, 1998; Cornet, 2007), or from experimental seeking of the maximum volumetric biomass productivity by varying the residence time in the PBR in continuous mode (see for the detailed explanations Takache *et al.*, 2009) and then finding the optimal biomass concentration. This ideal concentration must then be controlled by any mean, even if this seems easier to do in case of artificially-lightened continuous PBR (with constant incident LFD) rather than for solar one. This last case requires developing a control strategy in order to constraint the biomass concentration to oscillate around a pre-definite optimal set point, depending of the illumination conditions over the year.

## 2.2- Artificially-Lightened Photobioreactors

In case of artificially-lightened PBR, the main criterion is generally maximizing the biomass volumetric growth rate by an appropriate design, in agreement with the existing technologies for artificial light sources. In these conditions, the simple, reliable and general formula recently published (Cornet and Dussap, 2009) for the assessment of maximum (i.e. with $\gamma = 1$) volumetric performances of PBR may be practically used here with great benefit to analyze the best strategy of design. This useful relationship in fact gives the maximum volumetric biomass growth rate $<r_X>$ from three engineering variables, namely the incident LFD $\overline{q}_0$, the specific illuminated surface $a_{light}$ and the dark engineering fraction $f_d$. Using then the notations adopted in this paper enables to write it from the predictive knowledge of the energetic and quantum yields $\rho_M$ and $\overline{\psi}$, the half saturation constant for photosynthesis $K$ and the radiative characteristics of the medium ($\alpha$ for the radiative properties of the micro-organisms and $k_h$ for the angular distribution of the intensities as explained in appendix B) as follows (Cornet and Dussap, 2009):

$$<r_X>_{max} \equiv (1-f_d)\rho_M \overline{\psi} \frac{2\alpha}{1+\alpha} a_{light} \frac{K}{k_h} \ln\left[1+\frac{k_h \overline{q}_0}{K}\right] \quad (1)$$

The direct reading of this formula clearly establish that designing efficient PBR relies on technologies with low dark volume engineering fraction $f_d$ (the volume fraction of the reactor which is not lightened by construction), high values of the incident LFD $\overline{q}_0$ and mainly, high specific illuminated area $a_{light}$. For some classical designs of externally-lightened PBR, this specific area takes the following forms:

- rectangular PBR (thickness $L$) illuminated by one side: $a_{light} = 1/L$

- rectangular PBR (thickness $L$) illuminated by both sides: $a_{light} = 2/L$

- cylindrical PBR (diameter $D$) radially illuminated: $a_{light} = 4/D$     (2)

- cylindrical PBR with annular region (diameters $d$ and $D$): $a_{light} = \frac{4}{D}\frac{[1+d/D]}{[1-(d/D)^2]}$

This evidences that it becomes rapidly impossible to maintain constant the productivity of any geometry of reactor in the scale-up procedure because increasing significantly the characteristic length ($L$ or $D$) of the PBR to increase its volume, decreases in the same time the specific illuminated area $a_{light}$ and consequently the mean biomass volumetric growth rate as explained by eq. (1). Additionally, it appears also technically difficult to maintain the value of the incident LFD $\bar{q}_0$ in eq. (1) when increasing the illuminated surface for large size PBR, and in the same time, the requirement to have transparent walls on a large scale is often responsible of material stresses problems. Finally, because of the absorption of light by the cells and according to the previous analysis for optimal conditions, an increase of the characteristic size with a nearly constant LFD $\bar{q}_0$ (in the best situation), leads immediately to the appearance of an important dark volume inside the reactor ($\gamma < 1$) having then as a consequence a strong decrease of the biomass growth rate in comparison to maximum performances given by eq. (1). For all these reasons, it appears then almost impossible to scale-up the numerous artificially-lightened concepts already proposed in the literature at a scale of more or less 100 L without a proportional loss in biomass volumetric growth rate (eq. 1) leading roughly to a constant biomass dry weight production (in $kg_X$/h) at higher scale! This brief analysis simply demonstrates that externally-lightened PBR are not up-scalable, except by multiplying (as it is generally done) the number of small size units (around 100 L).

At the opposite, the only way to have easily up-scalable artificially-lightened PBR consists in developing internally-lightened PBR with internal structures generating light in situ (like fluorescent tubes – Muller-Feuga *et al*., 1998; Suh and Lee, 2003) or diffusing light from a point source (like diffusing optical fibers or plates – Janssen *et al*., 2003; Ogbonna *et al*., 1999). Additionally, it is noteworthy that this kind of concept can theoretically conserve the same mean biomass growth rate $<r_X>$, using a geometric similitude approach in the scale-up procedure, confirming then the strong theoretical potential of internally-lightened PBR in comparison to externally-lightened PBR. Their optimal design, requiring to find the best assembly of lighting structures inside the reactor volume, corresponds in fact exactly to the class of volumetrically-lightened reactors that we want to investigate in this article and that we have named PBR-HPV (for the French acronym "Haute Productivité Volumique"). Effectively, such a design enables to develop very important specific illuminated areas $a_{light}$ with also relatively high incident LFD $\bar{q}_0$ for large size of PBR (if the technological problems are solved), and to work theoretically with an optimal condition $\gamma = 1$ which will even become a criterion in designing the first construct of the reactor as explained later.

## 2.3- Solar Photobioreactors

Considering now the case of outdoor solar PBR leads first to a different approach because here, the main criterion is rather to maximize the surface biomass growth rate $<s_X>$ (relative to the sun collecting surface), and only in a second stage, to increase volumetric biomass productivities $<r_X>$ in order to minimize all the energetic costs associated with mixing and flowing of the culture medium. It is again possible to demonstrate that this former objective can be analyzed from a very simple formula for $<s_X>$, in the same manner that we have proceeded for $<r_X>$, but using an energetic basis instead of a kinetic approach. The rigorous expression of the thermodynamic efficiency of a PBR has been demonstrated a long time ago by the author from the theoretical or experimental knowledge of the mean volumetric biomass growth rate $<r_X>$ and the mean volumetric radiant light power density absorbed $<\mathcal{A}>$ inside the reactor (PAR). Considering only a global stoichiometric equation for the photosynthetic formation of one C-mole of biomass (Cornet *et al.*, 1998) with mean growth rate $<r_X>$ enables to write this exergetic yield in the form (Cornet *et al.*, 1994):

$$\eta_{th} = \frac{\sum_{p=1}^{Np} \upsilon_{pX} \frac{<r_X>}{M_X} \tilde{\mu}_p}{<\mathcal{A}> - \sum_{s=1}^{Ns} \upsilon_{sX} \frac{<r_X>}{M_X} \tilde{\mu}_s} \quad (3)$$

in which the $\tilde{\mu}_{s,p}$ are the chemical potentials for the substrates *s* and the products *p* involved in the biochemical reaction. This equation may be further simplified, assuming first that the arbitrary datum level (reference state) is chosen adequately and sufficiently close to the actual conditions for the calculation of the Gibbs energies of formation, and second, demonstrating from its calculation (Roels, 1983) that the chemical power density term for the substrates appears negligible in front of the radiant light power density, leading then to the more simple relation:

$$\eta_{th} \cong \frac{<r_X> \Delta g_X'^0}{M_X <\mathcal{A}>} \quad (4)$$

Here again, it is possible to adapt this simple equation to our case of interest, i.e. the situation giving the maximum volumetric (or surface) kinetic performances using the requirement $\gamma = 1$ as previously seen for the PBR functioning. Then, the maximum volumetric biomass growth rate in the PBR $<r_X>_{max} = <s_X>_{max} a_{light}$ is obtained when the incident LFD $\bar{q}_0$ is entirely absorbed in the reactor verifying $<\mathcal{A}>_{max} \cong a_{light} \bar{q}_0$ (Cornet and Dussap, 2009; Pruvost *et al.*, 2008) and giving finally the sought formula for the maximum surface growth rate from (4):

$$<s_X>_{max} \cong \frac{\eta_{th} M_X \bar{q}_0}{\Delta g_X'^0} \quad (5)$$

Considering that the standard Gibbs enthalpy of formation for the biomass $\Delta g_X'^0$ is nearly constant in most of the operating conditions (Cornet *et al.*, 1998; Roels, 1983), this useful equation demonstrates that the maximum surface biomass growth rate (fulfilling the optimal condition $\gamma = 1$) is only a function of the incident LFD $\bar{q}_0$. This point is confirmed by the Figure 2 on which experimental results of PBR of different geometries cultivating the cyanobacteria *Arthrospira platensis* (with nitrate as nitrogen source), but all operating in optimal condition ($\gamma = 1$) have been used to calculate the thermodynamic efficiency of the reactor $\eta_{th}$ as a function of the incident LFD $\bar{q}_0$ (from eq. 4). These results quantitatively confirm on a strong physical basis that the thermodynamic efficiency of the photosynthesis strongly decreases with increasing the incident LFD on the PBR with a power law approximately given by (Figure 2):

$$\eta_{th} \propto \bar{q}_0^{-\frac{1}{5}} \quad (6)$$

Finally, using eqs. (5-6) leads straightforwardly to the conclusion that the mean maximum surface biomass growth rate follows also a power law in the form:

$$<s_X>_{max} \propto \bar{q}_0^{0.8} \quad (7)$$

These simple equations (5-7) are of considerable interest for the field of PBR engineering because they have two strong involvements that can be summarized as follows:

(*i*) conversely to the thermal solar energy conversion processes, it is always inadequate to concentrate the sun on a PBR considering maximum surface performances (the increase in immobilized surface corresponding at least to the geometric concentration factor then proportional to the increase of $\bar{q}_0$);

(*ii*) for the concepts of PBR using a direct solar collecting surface through transparent walls as it is generally the case, the maximum surface productivities do not depend in any way on the design of the reactor (if the requirement $\gamma = 1$ is fulfilled), but they are a purely thermodynamic consequence only linked to the incident LFD $\bar{q}_0$ (eq. 5).

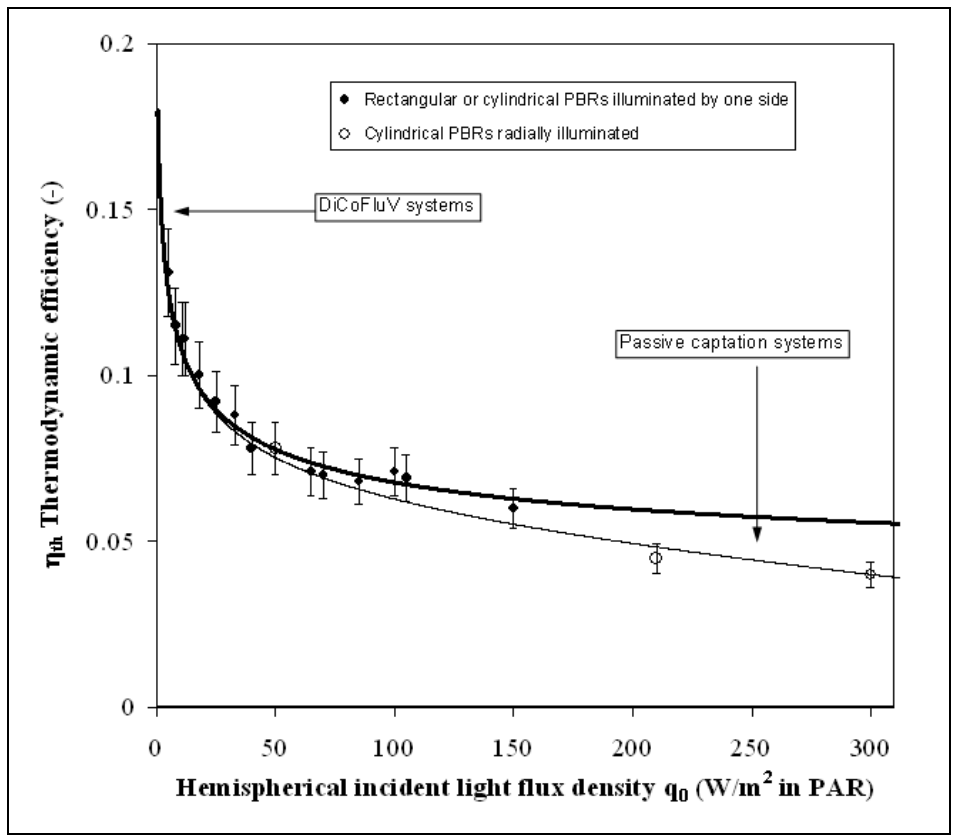

Figure 2: Thermodynamic efficiencies (or exergetic yields) of different kinds of photobioreactors versus the incident hemispherical light flux density $\bar{q}_0$. All the experiments reported on this figure have been obtained in optimal conditions with a working illuminated fraction $\gamma = 1$ in cultivating the cyanobacterium *Arthrospira platensis* on a conventional medium using nitrate as N-source. The respective values of mean LFD $\bar{q}_0$ for classical solar passive PBR or for the solar DiCoFluV concept PBR are indicated.

Thus, the only mean to increase the maximum surface productivities of a solar PBR consists in finding a solution in eq. (5) to uncouple the incident LFD $\bar{q}_0$ and the thermodynamic efficiency $\eta_{th}$ (obligatorily related by eq. 6 and Fig. 2 for direct solar collection processes), in order to maximize each of them independently. This conclusion is exactly at the origin of the DiCoFluV concept (for the French acronym "Dilution Contrôlée du Flux en Volume") in which we propose to capture the sun with an independent tracking system at its maximum hemispherical incident flux density $\bar{q}_\perp$ (see appendix A), to concentrate it in introducing radiation in lighting structures enabling to dilute this radiant power at a low incident LFD $\bar{q}_0$ (on the surface of the structures) inside the PBR volume, corresponding then to the highest possible thermodynamic efficiency. Effectively, for ideal solar conditions (see appendix A), the mean incident LFD $\bar{q}_0$ in the PAR is roughly $\bar{q}_0 = \bar{q}_\perp = 340 \text{ W/m}^2$ with a tracking system instead of $\bar{q}_0 = \bar{q}_\cap = 250 \text{ W/m}^2$ for direct passive solar collection, and in the same time, as it can be seen on Fig. 2, it is possible to use the radiation at a low incident LFD with an efficiency around 15% (relative to the PAR, cultivation on nitrates, Fig. 2), instead of an efficiency of around 5% (PAR) at $\bar{q}_0 = \bar{q}_\cap = 250 \text{ W/m}^2$ in case of direct passive collection. Using then eq. (5) easily demonstrates that these two factors enable increasing the mean surface maximum biomass growth rate by a factor 4, corresponding to a thermodynamic limit for photosynthesis efficiency (impossible to overreach by any other mean!) with this concept. This clearly establishes again, like for artificially-lightened PBR, the strong theoretical superiority of volumetrically- and internally-lightened concepts in solar conditions, on the basis of maximizing surface biomass productivities. In the same manner, the optimal design of such reactors requires investigating general methods to find the best assembly of lighting structures inside the culture volume, as proposed in the following of this article. In this DiCoFluV concept, the optimal operating value of the dilute incident LFD must be chosen as a compromise between a high enough value with regard to the irradiance of compensation (corresponding to a state where 50 % of the energy is supplied by the respiration) and a low enough value maintaining a high thermodynamic efficiency (Fig. 2). For cyanobacteria involved in the results appearing on Figure 2, this compromise leads to dilute the incident optimal LFD down to $\bar{q}_0 \cong 3-5 \text{ W/m}^2$.

Alternatively, examination of eq. (1) again shows that working with a very low (dilute) incident LFD $\bar{q}_0$ to increase the thermodynamic efficiency and then the surface kinetic performances will clearly strongly decrease (with a logarithmic law) the volumetric biomass productivities. Fortunately, in the DiCoFluV concept, the necessity to develop a high internal lightened area imposes also to increase significantly the specific illuminated area $a_{light}$ enabling to compensate for (as proved by eq. 1), in a given extend, losses due to the low LFD value. Generally speaking, it remains however difficult to maximize together surface and volumetric kinetic performances and, as proposed above, the choice of a main criterion is necessary to properly optimize the design of a given PBR.

# 3- Optimal Assembly of the Lighting Structures inside the Photobioreactor Using the Constructal Approach

The previous analysis has clearly established the very high potential interest of volumetrically and internally-lightened concepts, both for artificial or solar large-size PBR. It will even be demonstrated in the following that they correspond for most optimal criteria to ideal concepts of PBR. This class of reactors, artificially-lightened or solar, has an important common design characteristic consisting in finding the optimal assembly of the lighting structures (of any given shape) inside the reactor culture volume (also of any given shape, but generally rectangular or cylindrical). This is a similar problem except, as explained above, in the choice of the constraints which are generally different for artificial (maximization of the volumetric biomass growth rate) or solar (maximization of the surface biomass growth rate first) illuminations in the optimization procedure. Such optimization of geometric designs may be envisaged in the framework of the recent constructal approach developed by Bejan (2000; www.constructal.org), which is particularly well-suited for engineering purposes as envisaged in this article.

In the following, we will privilege mainly cylindrical geometries - i.e. examining extensively the two-dimensional problem of finding (in the cross section of the PBR) the best array of small tubular lighting structures in a cylindrical reactor - for evident reasons. First, the cylindrical shape for the reactor itself is a classical design for bioreactors and most of the air-lift or bubble column systems (the pneumatic option for mixing is an obligation when putting lighting structures inside the reactor and when low dissipated power is required) and such systems are industrially up-scalable to hundreds of cubic meters. Second, the cylindrical shape for the lighting structures is also the more representative of the main technologies involved for artificial illumination (fluorescent tubes, light electroluminescent threads…) or solar light diffusion (optical fibers) with numerous different commercial sizes available. This appears then as the more realistic technical option to develop first optimized pilot plants of volumetrically-lightened PBR as it is currently investigated in our laboratory (www.biosolis.org).

Nevertheless, the general geometrical optimization problem, dealing also with rectangular and spherical geometries will be briefly sketched in a following paragraph to compare the strengths and drawbacks of each concept.

## 3.1- The Optimal Assembly for the 2D-cylindrical Problem

From our previous choices, the general geometrical optimization problem is greatly oversimplified because working in continuous condition with (in a first approximation) a completely stirred tank reactor imposes a constant value for the biomass concentration inside the reactor volume. Additionally, we will assume in the following that the incident LFD on the lighting structures is homogeneous on the entire length of the structure. It is a reasonable assumption because for artificial illumination, fluorescent tubes for example generate really homogeneous LFD (Cassano *et al*., 1995), and it is technically possible to develop superficial treatment of optical fibers to obtain also lateral homogeneous LFD all along the fiber.

The constructal approach relies on simple general principles (in fact statements inspired from the so-called variational principles of physics), the first of which states that (Bejan, 2000) "*For a finite-size system to persist in time, its configuration must evolve such that it provides easier and easier access to its currents.*" In the perspective of finding optimal designs for steady-state processes, this sentence imposes to seek solutions which maximize the access to the flux, or identically, which minimize the resistance to the flux feeding and maintaining the structure of the considered dissipative system. A consequence of this principle is that the imperfections are then distributed in such a way that they generate the shape and structure of the system. This emphasizes that the determinism has a privileged direction and that any design must be optimized step by step, from the smallest scale to the final one (the application), distributing the imperfections in putting the more resistive regime at the smallest scale of the system. In our case study, this clearly corresponds to the scale at which the photon transport process operates, simultaneously with the absorption by the material phase where the photosensitized reaction occurs. In the 2D-cylindrical geometry investigated, this corresponds indeed to the distance $d_i$ between the lighting structures (see Fig. 3), which must be optimized in a first construct. With our previous assumptions of homogeneity in biomass concentration and incident LFD, the second construct suit then with the scale of the whole reactor, enabling to calculate the distance $d_S$ for the lighting structures (Fig. 3), directly deduced from an optimal surface-to-volume optimization, and imposing finally the number of structures for a given diameter of the PBR.

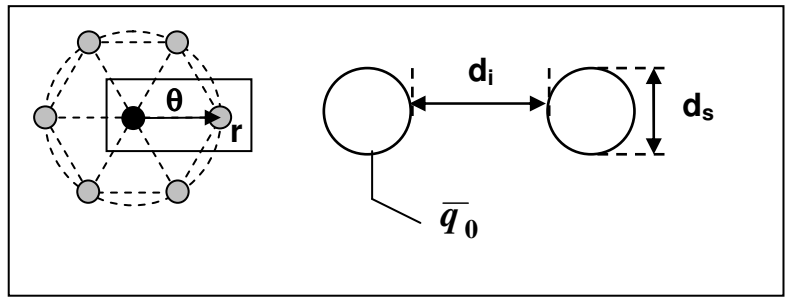

Figure 3: Schematic representation and key variables for the optimal assembly problem using the constructal approach in case of a 2D-cylindrical geometry. The system of coordinates used for the mathematical treatment is represented in case of a triangular grid (more compact). The two characteristic dimensions of the problem, the distance between lighting structures (with incident LFD $\bar{q}_0$ ) $d_i$, and the structures diameter $d_S$ are indicated and defined.

### *3.1.1- Finding the optimal distance $d_i$ at the smallest scale – The first construct*

The previously sketched geometrical optimization problem is presented on Figure 3 showing the arrangement of lighting structures in a triangular grid (the more compact one), with the two characteristic distances to optimize $d_i$ and $d_S$. The first construct consists in finding the optimal distance between structures $(d_i)_{opt}$, as the smallest physical scale of the system. This is a typical radiative transfer problem requiring to solve the radiative transfer equation (RTE) with the associated optimal constraint (at this scale) that $\gamma = 1$ (see above for explanations). This problem can be handled with sophisticated numerical methods for solving the RTE in the two-dimensional cylindrical geometry $(r, \theta)$, but it is also possible to formulate some adequate assumptions in order to degenerate the problem to a quasi-one-dimensional treatment, enabling to use the general analytical solutions of the RTE developed in appendix B of this article. Considering that we seek for our application the radiation field in the limiting case (the worst case regarding the $\theta$ angular dependence of the field around each structure) where the compensation point for photosynthesis (low irradiance value $G_C$) is reached between two respective structures (at a distance $d_i/2$ corresponding strictly to the local constraint $\gamma = 1$), and then neglecting in this condition the photons incoming from others structures in the vicinity (strong light limitation with high biomass concentration), corresponds to find the radiation field only between two structures in cylindrical geometry, assumed to be independent from the others (see the selected faced to faced structures for example on Fig. 3).

In this particular situation, and assuming a given quasi-homogeneous incident hemispherical light flux density (LFD) on each lighting structure $\bar{q}_0$ in the PAR (the boundary conditions for $I^\pm$) enables to solve the general eq. B7 (appendix B) with $m = 1$ for a cylindrical geometry (see appendix B for detailed explanations). The field of irradiance (spectrally-averaged on the PAR then omitting the $\lambda$ indices) between the two structures on the basis of a frame of reference defined on Fig. 3, with $c_1 = r$, is then given by:

$$\frac{G}{\bar{q}_0} = 2 k_h \frac{K_1(\delta r_T) I_0(\delta r) + I_1(\delta r_T) K_0(\delta r)}{[K_1(\delta r_T) I_0(\delta r_S) + I_1(\delta r_T) K_0(\delta r_S)] + \alpha [K_1(\delta r_S) I_1(\delta r_T) + I_1(\delta r_S) K_1(\delta r_T)]} \quad (8)$$

in which $I_n(x), K_n(x)$ are respectively the first and second kind modified Bessel functions of order $n$, and with the following definitions of the parameters (see Fig. 3):

$$r_T = \frac{d_i + d_s}{2} \qquad r_S = \frac{d_s}{2}$$

$$\alpha = \sqrt{\frac{a}{a + 2\bar{b}s}} \qquad \delta = k_h \sqrt{a(a + 2\bar{b}s)} \quad (9)$$

$$a = Ea\, C_X \qquad s = Es\, C_X$$

This equation thus provides a very good approximation for the radiation field between two lighting structures, taking into account with the degree of collimation $n$ the angular distribution of the radiation (depending mainly of the model of emission for the sources) in the more subtle way than it is possible to do with the two-flux principle for the RTE solution (see appendix B). This point is of crucial importance because it has strong influence to the irradiance field calculations $G(\mathbf{r})$ and then to the volumetric radiant power density absorbed $\mathcal{A}(\mathbf{r})$ enabling the kinetic coupling formulation (Cornet and Dussap, 2009).

Equations (8-9) have been used to draw Figure 4 from the knowledge of the compensation point for photosynthesis $G_C$ (the well-known case of *Arthrospira platensis* is here practically considered), leading, in the framework of the constructal approach, to the calculation of the optimal distance between structures $(d_i)_{opt}$ versus the incident LFD $\bar{q}_0$, the biomass concentration $C_X$ appearing as parameter. This clearly demonstrates that, in the absence of additional constraint, this optimization procedure presents two degrees of freedom (DOF). As we have discussed above, the incident LFD $\bar{q}_0$ first is not really a DOF for the DiCoFluV concept, because it requires working at low values of $\bar{q}_0$ to ensure a maximum energetic efficiency for the photosensitized reactions in the culture medium, and then its value is fixed for a given micro-organism. Conversely, for the HPV concept, using artificially-lighting structures, the value must be chosen as higher as possible (see eq. 1) in order to increase the volumetric performance of the PBR, so it becomes generally a technical limitation for each considered technology of involved sources. On the other hand, the biomass concentration $C_X$ remains generally a DOF of the problem and, as it can be seen on Fig. 4, an increase of $C_X$ decreases the optimal distance $(d_i)_{opt}$ contributing then to increase the specific illuminated area in the PBR and thus to obtain higher volumetric productivities. Nevertheless, it probably exists an hydrodynamic limit for this parameter when $(d_i)_{opt}$ is around one or two millimeters with $C_X$ reaching roughly 30-50 g/L and besides, a high biomass concentration in the reactor increases the risk of adhesion on the lighting structures. In any case, the maximum value of $C_X$ must be chosen a priori and experimentally verified a posteriori by adjusting the residence time in the PBR from the calculated maximum biomass productivity issued from the complete optimization procedure (including the following step for $d_S$).

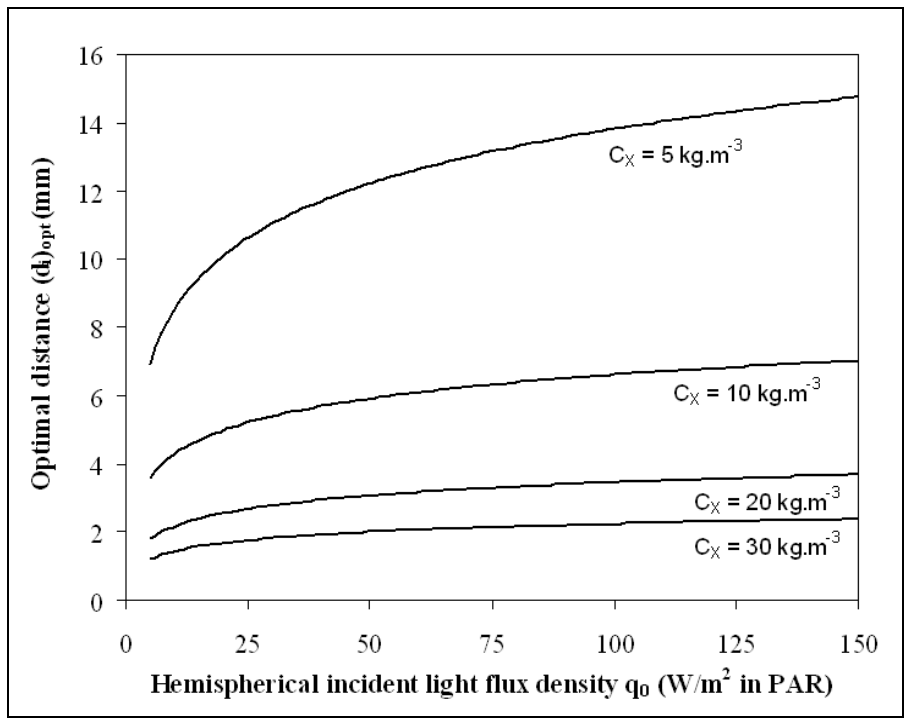

Figure 4: Optimal distance between lighting structures $(d_i)_{opt}$ versus the hemispherical incident light flux density $\overline{q}_0$ resulting from the optimization at the smallest scale (first construct of the constructal approach) and calculated on the basis of appendix B, using the constraint $\gamma = 1$. The biomass dry weight concentration $C_X$ is indicated as parameter. For all calculations, the radiative properties of *Arthrospira platensis* have been used.

## *3.1.2- Finding the optimal distance $d_S$ at the largest scale – The second construct*

Once the smallest scale is resolved, the second level must be optimized. It has been already discussed that it corresponds for our problem to the calculation of the optimal diameter of the lighting structures $(d_S)_{opt}$, involving then to maximize the mean volumetric biomass growth rate $<r_X>$ (taken as a function of $d_S$) at the scale of the whole PBR. This theoretically requires formulating the local kinetic coupling with the radiation field using energetic and quantum yields (Cornet and Dussap, 2009), writing $<r_X>$ in the form:

$$<r_X> = \frac{1}{V}\iiint_V r_X[G(\mathbf{r}),d_s]\Big|_{C_X,\bar{q}_0,(d_i)_{opt}} dV \quad (10)$$

before maximization. This tedious numerical procedure can be however avoided using the very good approximation of $<r_X>$ given by eq. (1) and demonstrating that this problem becomes a purely geometric surface-to-volume optimization. This relationship may be rewritten here for the particular application of an internally-lightened PBR, the specific area $a_{light}$ becoming the product of a specific illuminated area for the structures $a_S = 4/d_S$ and its corresponding total volume fraction (relative to both the culture broth and the internal lighting structures volumes) $\varepsilon = \varepsilon_{max}/[1+d_i/d_S]^2$. Using the well-known value for $\varepsilon_{max}$ corresponding to the situation where each lighting structure touches each other leads a simple expression for $a_{light}$ as function of the structures diameter $d_S$:

$$a_{light}(d_S) = a_S \varepsilon = \frac{4}{d_S}\frac{\varepsilon_{max}}{\left[1+\dfrac{d_i}{d_S}\right]^2} = \frac{4}{d_S}\frac{\pi/(2\sqrt{3})}{\left[1+\dfrac{d_i}{d_S}\right]^2} \quad (11)$$

Considering now that in eq. (1) giving $<r_X>$ anything is kept constant except the design variables (the domain sensitivity is in fact examined) appearing in eq. (11), and that $d_i$ has been optimized at the preceding step (the first construct), it is finally easy to seek the optimal value of the lighting structures diameter $d_S$ from the criterion:

$$<r_X>_{max} \Rightarrow \left(\frac{\partial a_{light}}{\partial d_S}\right)\bigg|_{C_X,\bar{q}_0,(d_i)_{opt}} = 0 \quad (12)$$

giving the simple straightforward result:

$$(d_S)_{opt} = d_i = (d_i)_{opt} \quad (13)$$

which leads to the optimal value of the volume fraction for the lighting structures inside the volumetrically-lightened PBR (neglecting the edge effects):

$$\varepsilon_{opt} = \frac{\varepsilon_{max}}{4} = \frac{\pi}{8\sqrt{3}} = 0.2267 \quad (14)$$

These results complete the optimal design calculation for the particular case of a 2D-cylindrical concept. The constructal approach that we have used to find this result guarantees in fact that, if the constraints and the objectives have been properly defined in their mathematical formulation, then we obtain the best geometrical design, i.e. the design providing the optimal distribution of imperfections (the irreversibilities) for the volumetric dissipative radiant light transfer problem in a reactive photosensitized media.

As an illustration of the constructal optimized design for a 2D-cylindrical light transfer problem, the best array of lighting structures is represented on Figure 5 for two different cases leading to almost the same volumetric performances of a 220 mm inner diameter PBR. The first situation (Fig. 5a) corresponds rather to an artificially-lightened PBR-HPV with for example, fluorescent tubes of 12 mm delivering an incident LFD $\bar{q}_0 \cong 100$ W/m² and operating with a biomass concentration $C_X = 6$ g/L (($d_i$)$_{opt}$ = 12 mm) leading to assembly 70 lighting structures. The second situation (Fig 5b) corresponds to a solar PBR-DiCoFluV concept using lateral diffusing optical fibers of 2 mm to dilute the sunlight radiation to $\bar{q}_0 \cong 5$ W/m² (in the range of the optimal value for cyanobacteria as explained above), with a biomass concentration $C_X = 18$ g/L (($d_i$)$_{opt}$ = 2 mm) requiring to assembly 2640 lighting structures inside the reactor volume. However, as it can be seen on Fig. 5c, the same maximum volumetric mean biomass growth rate are obtained in these two different examples at a value of $d_S$ satisfying indeed eq. (13), and corresponding (Fig. 5d) to the same volume fraction given by eq. (14). This demonstrates that, as explained above, it is possible to reach the same volumetric performances with a DiCoFluV concept imposing a very low value for the incident LFD $\bar{q}_0$ if a high specific illuminated area is obtained by using small diameters $d_S$ for the lighting structures. This increase of specific illuminated area is of course related to a necessary increase in the biomass concentration at which the continuous PBR operates.

Additionally to the optimization procedure obtained in this paragraph, it must be pointed out that the very simple results obtained at the PBR scale rely on the previous assumptions of constant biomass concentration $C_X$ and constant incident LFD along the lighting structures, considered here as DOF. If one of these conditions was not respected (for example tubular PBR with no recycle for $C_X$ or a non-treated coated diffusing optical fiber with variable $\bar{q}_0$), it would become a constraint of the problem and it would be necessary to optimize a new intermediate construct, leading to the so-called tree shape for the assembly of lighting structures as it is often the case in many constructal applications (Bejan, 2000). Finally, regarding now the thermodynamic significance of the previous optimization approach enables to analyze the status of the constructal law applied to a radiative transfer problem since the maximization of the light flow rate access is ensured by controlling the local absorbed radiant light power density field, or the convergence of the radiant light flux density (Cornet, 2005), inside the reactive material phase of the PBR. From the expression giving the entropy production rate in a PBR (Cornet *et al.*, 1994), it is then possible to demonstrate (but out of the scope of this paper) that the best assembly resulting from the constructal optimization is equivalent, for this problem, to a relative maximization of the exergetic yield of the PBR at a given LFD $\bar{q}_0$. Nevertheless, because it works with optimal and low incident LFD $\bar{q}_0$, only the DiCoFluV concept obeys to the general variational law, corresponding to an absolute maximization of the exergetic yield for the whole PBR (see Fig. 2):

$$\mathrm{d}\eta_{th} = 0 \quad (15)$$

This result demonstrates rigorously why this concept can be considered as a thermodynamic limit of the sunlight conversion process in a PBR and contributes to clarify the thermodynamic statement of the constructal law for this particular engineering application.

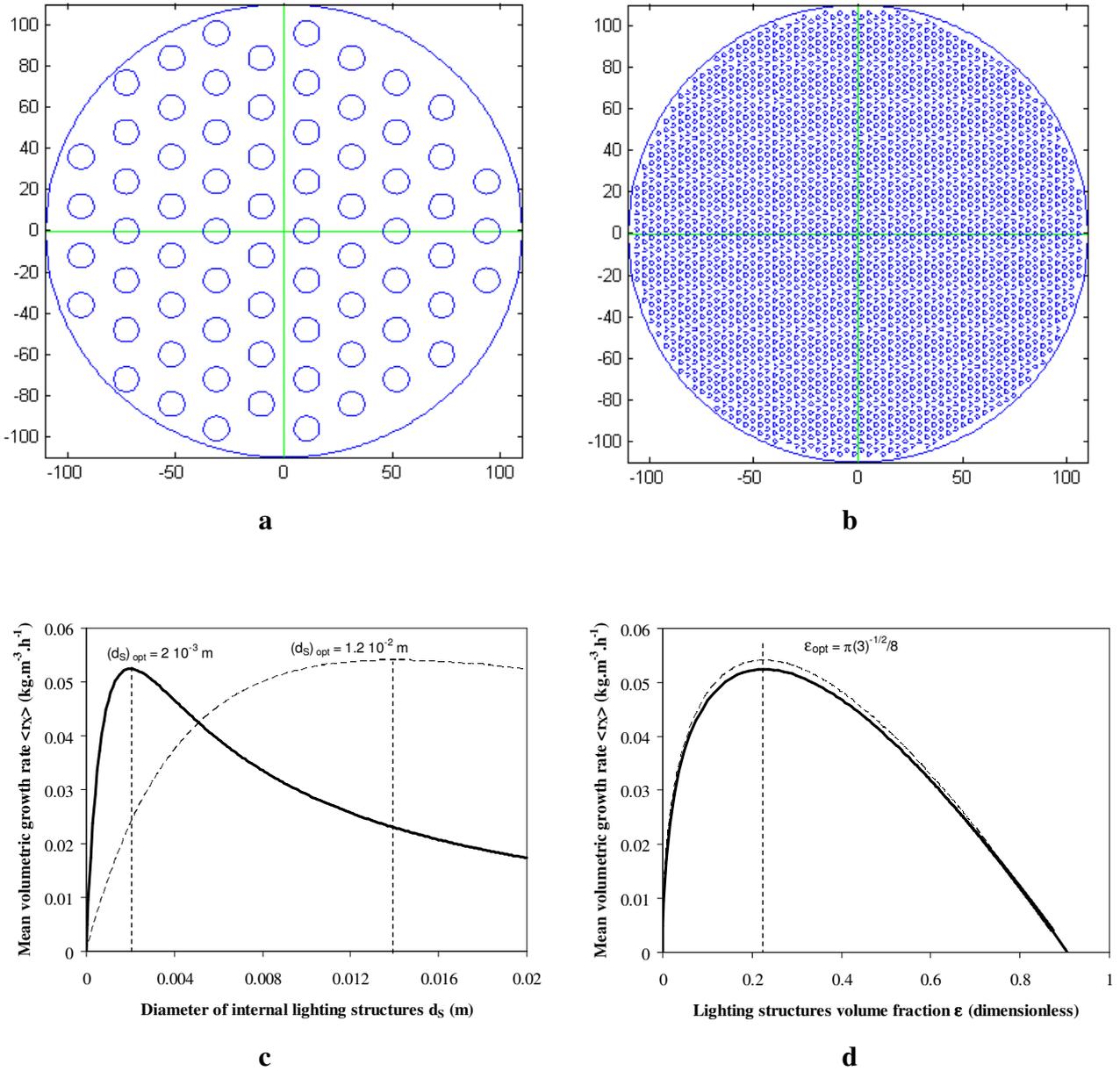

Figure 5: Optimal assemblies obtained for two different kinds of PBR (total diameter $D$ = 220 mm) verifying for the 2D-cylindrical geometry the optimal value of the second construct $(d_S)_{opt} = (d_i)_{opt}$. The first kind of PBR (*a*) represents rather an artificial illumination HPV concept with an incident LFD $\bar{q}_0$ = 100 W/m² and with lighting structure diameters $(d_S)_{opt}$ = 12 mm, obtained for an ideal biomass concentration $C_X$ = 6 g/L (dotted lines on figures *c* and *d*). The second kind of PBR (*b*) represents a solar illumination DiCoFluV concept with an incident LFD $\bar{q}_0$ = 4-5 W/m², with lighting structure diameters $(d_S)_{opt}$ = 2 mm, obtained for an ideal biomass concentration $C_X$ = 18 g/L (continuous lines on figures *c* and *d*).

These values have been chosen to give roughly the same volumetric biomass productivity $<r_X>$ in both cases as demonstrated from two different representations, using in abscissa the diameter of internal lighting structures $d_S$ (**c**) and then confirming the optimal values $(d_S)_{opt}$ obtained by constructal optimization, or using the lighting structures volume fraction $\varepsilon$ (**d**) confirming also the common optimal value at $\varepsilon_{opt} = \pi/(8\sqrt{3})$.

## 3.2- The Optimal Assembly for Other 1D and 3D Geometries

The methodology presented above to find the optimal assembly of lighting structures inside volumetrically-lightened PBR is of course not limited to the case of 2D-cylindrical geometry that we have decided to develop as the more practical case study. In fact, the proposed approach may be generalized to other geometries and coordinates systems corresponding to also possible concepts and designs of volumetrically-lightened PBR. The 1D-Cartesian geometry for example corresponds to the assembly of lighting panels in rectangular multi-layered PBR (Ono and Cuello, 2006; Zijffers et al., 2008) and the 3D-spherical geometry could represent concepts of PBR internally-illuminated by luminescent particles.

The generalization of the previous constructal approach developed for cylindrical geometries requires first to use a general theoretical tool enabling to calculate the optimal distance $d_i$ for the first construct. For this, the general case of quasi-one-dimensional two-flux analytical solutions of the RTE has been in fact already developed in appendix B in any geometry and may then be used as a rigorous approach for rectangular coordinates or as a first good approximation in curvilinear coordinates (see the establishment of eq. 8). The second step requires extending the methodology used to find the optimal size of lighting structures $d_S$ to any geometry from eq. (1). This work has been done and reported in appendix C of this article enabling to focus here the discussion to the obtained results of the optimization procedure. They give the optimal size of the considered lighting structures $(d_S)_{opt}$ for the second construct in any geometry, once the first and more resistive scale has been resolved for $(d_i)_{opt}$. The resulting dimensionless mean volumetric biomass growth rates (relative to the maximum 1D-rectangular case) have been plotted versus the dimensionless abscissa $d_S / d_i$ on Figure 6. It appears that the maximum performances are effectively reached at the respective value of $(d_S)_{opt}$ obtained by the constructal optimization and given by eq. (C5) in appendix C for each considered geometry. These maximum volumetric kinetic performances demonstrate that the optimal assembly for 2D-cylindrical geometry is 40% more efficient than the corresponding 3D-spherical geometry, but for characteristic dimensions $d_S$ two time smaller. Finally, using 1D-rectangular systems with smaller and smaller characteristic thicknesses $d_S$ can lead to a significant increase of the kinetic performances of the PBR if mixing and hydrodynamic problems can be solved in this geometry. At the opposite, this efficiency rapidly decreases with increasing the thickness of the lighting structures. Fig. 6 finally appears as an interesting tool enabling to discuss in a concise manner the underlying consequences of a given choice of geometry, in connection to other constraints for optimal PBR engineering design.

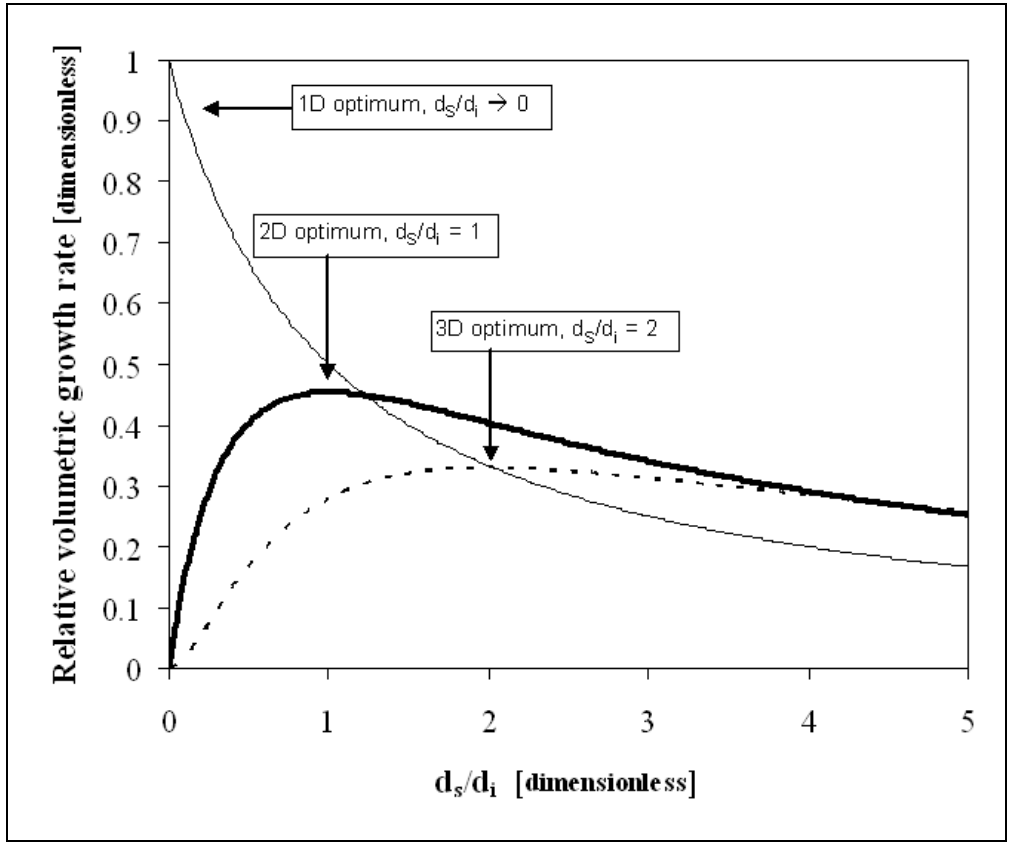

Figure 6: Mean relative volumetric biomass growth rates (normalized respectively to the maximum value obtained in the 1D-rectangular geometry for $d_S/d_i \to 0$) versus the dimensionless abscissa $d_S / d_i$ (in which the distance between structures $d_i$ is fixed to its ideal value $(d_i)_{opt}$) for the three geometries examined in the article (rectangular, cylindrical and spherical). In each case, the optimal value obtained for biomass productivities confirms indeed the calculated optimum given by eq. (C5) for the second construct with the corresponding definition of *m* (see appendix B, scheme B1).

( —— ) 1D-rectangular geometry; ( ▬▬ ) 2D-cylindrical geometry; ( - - - - ) 3D-spherical geometry

## 4- The Concept of Ideal Photobioreactor

The concept of ideal reactor (in term of flow but mainly in term of design) is a classical approach in the field of chemical engineering (Aris, 1999) that we intend to generalize here to the case of PBR. We have already developed in fact the practical relations useful to tackle this problem in the first part of this article dealing with optimal constraints and design. Defining indeed an ideal PBR requires to demonstrate and calculate the concept giving the best kinetic performances as a theoretical limit, independently of any technical or implementation considerations. As it has been previously discussed, the methodology appears different according as surface or volume maximum biomass growth rates are investigated.

### 4.1- Ideal Surface Productivities

This first problem is really a case study because eq. (5) has already demonstrated that the maximum surface mean biomass growth rate, mainly interesting for solar applications, can serve as a definition of the ideal outdoor solar PBR. It gives indeed actually the maximum performances as a function of only purely thermodynamic data like the exergetic yield of the PBR $\eta_{th}$ considered at its maximum value of around 6% on ammonia as substrate (i.e. 25% higher than with nitrate; see Cornet, 2007; Cornet and Dussap, 2009) for a mean daily ideal value of the LFD $\bar{q}_0 = \bar{q}_\cap = 250 \text{ W/m}^2$ (passive solar collection through transparent walls, see appendix A) and the Gibbs free energy for the biomass $\Delta g_X^{\prime 0}$, roughly taken at $6 \times 10^5$ J/mol for a mean photosynthetic micro-organism (Cornet, 2007; Roels, 1983). Hence, the result appears completely independent of the design and for any concept of direct passive solar collection, taking a mean C-molar mass at $2.4 \times 10^{-2}$ kg$_X$/C-mol$_X$ (Cornet and Dussap, 2009) and considering a mean yearly sunlight duration $t_S$ (see appendix A) leads finally (for a light-limited PBR only) from eq. (5) to:

$$<s_X>_{max} \cong 100 \text{ t}_X.\text{ha}^{-1}.\text{y}^{-1} \quad (16)$$

In the same manner, considering now the possibility to dilute the radiation as proposed in the DiCoFluV concept, together with a higher uncoupled LFD using a tracking system of solar collection giving roughly $\bar{q}_0 = \bar{q}_\perp = 340 \text{ W/m}^2$ (see appendix A) enables to work at a maximum exergetic yield of about 17% (with ammonia as N source; see Fig. 2 and Cornet, 2007) leading then from the same formula to (with an optimal dilution factor of about $C = 1/70$):

$$<s_X>_{max} \leq 400 \text{ t}_X.\text{ha}^{-1}.\text{y}^{-1} \quad (17)$$

These interesting results are of crucial importance for outdoor solar PBR engineering because first, they give, on strong physical and energetic bases, the maximum surface productivities that it is possible to reach with ideal reactor and sunlight conditions over the year, and second they demonstrate that the DiCoFluV concept represents exactly the ideal PBR in this case (i.e. any theoretical DiCoFluV concept of PBR is an ideal reactor regarding surface productivities), corresponding then to a thermodynamic limit in using photosynthetic reactive systems to satisfy human needs.

Obviously, using one of these concepts for sunlight to biomass conversion at any location on earth, with actual sunlight illumination conditions, requires correcting these values, then leading to significantly lower surface performances.

## 4.2- Ideal Volumetric Productivities

Conversely to surface kinetic performances, volumetric performances can be assessed from eq. (1), demonstrating that in any case, it appears difficult to avoid any technological description of the PBR because it is necessary to evaluate the specific illuminated area $a_{light}$ and the incident LFD, which may be a technical factor for artificially-lightened PBR. It is however possible to define a limiting value for the specific illuminated area from purely theoretical radiative transfer considerations. Considering indeed the mean radiative properties for photosynthetic micro-organisms (Berberoglu *et al.*, 2008; Cornet and Dussap, 2009; Pottier *et al.*, 2005; Takache *et al.*, 2009) and using the RTE to calculate the ideal ($\gamma = 1$) minimum length of incident LFD attenuation (considering the mean value of $\bar{q}_0 = \bar{q}_\cap = 250 \text{ W/m}^2$ corresponding to passive sunlight collection as the highest reference value) leads roughly to a thickness of 100 µm (in the sense of a skin thickness for electromagnetic applications) as a weak function of the characteristic size of the considered micro-organism. Using then eq. (1) with this value of $a_{light} = 10^4$ m$^{-1}$ and the previous incident LFD together with the values of other parameters available in the work of Cornet and Dussap (2009) gives then the ideal volumetric biomass growth rate (ideal solar PBR with ammonia as N source):

$$<r_X>_{max} \cong 70 \text{ t}_X.\text{m}^{-3}.\text{y}^{-1} \quad (18)$$

For artificially-lightened PBR, the ideal value is likely of the same order of magnitude, using lower value for the incident LFD (because of a technical limitation) but 24 hours a day of illumination. Unfortunately, this limiting and ideal value corresponds to a completely conceptual approach because the previous very low optical thickness has been calculated considering that each cell touches each other in a strong attenuating thin biofilm. In these conditions indeed, the biomass concentration reaches 200-250 g/L and it is not possible to manage the hydrodynamic behavior of the PBR. The true optimal thickness for ideal light attenuation compatible with a correct hydrodynamics of the culture broth is rather probably around 2 mm, i.e. giving volumetric performances 20 times lower than those of eq. (18). It must be noticed again that we refer ourselves here to only quasi-ideal volumetric productivities because they are never independent of some technological considerations. These quasi-ideal values can probably be reached with direct optimized outdoor sunlight collector systems (Doucha and Livansky, 2006; www.biosolis.com), or alternatively with artificially and volumetrically-lightened concepts developed in this article like PBR-HPV. Nevertheless, as demonstrated by Fig. 5, the low incident LFD imposed for the DiCoFluV concept always leads to lower biomass productivities (even with high specific illuminated areas), showing that this is not the ideal concept in this case and that it is impossible to optimize together surface and volumetric kinetic performances because of the opposite requirements for the incident LFD $\bar{q}_0$.

## 5- Some Technological Remarks about the DiCoFluV Concept

In all the previous part of this article, the DiCoFluV concept of PBR has been defined as the ideal reactor maximizing the surface biomass productivities by using internal lighting structures enabling to dilute the incident solar light flux density at the highest thermodynamic efficiency available for photosensitized reactions engineering. In many practical cases however, the presence of lighting structures directly in the culture medium of the PBR may be considered as a technical problem regarding mixing and hydrodynamics of the liquid phase and mainly, regarding the possible adhesion of the micro-organisms on these structures. Fortunately, this concept is not limited to internally-radiated PBR and can be easily extended to externally-radiated DiCoFluV PBR as showed on Figure 7. In this last case, the lighting structures are alternatively disposed between culture chambers of any geometry (mainly rectangular, cylindrical or tubular) to ensure the dilution of the incident solar light flux which is transmitted in a classical manner, through transparent walls of the PBR (see some examples on Fig. 7). In these conditions indeed, the optimal assembly and sizes of lighting structures and culture chambers may be obtained using the constructal approach, in a similar way as proposed along with this article but with additional complexities in the light transfer problem requiring generally using the geometrical optics to treat the interfaces. For all these externally-radiated concepts of PBR however, the volumetric kinetic performances will be lower because these systems are always less compact than the corresponding internally-radiated one, but importantly, it must be emphasized that the surface productivities (corresponding to the major limiting step in any solar conversion process) will be theoretically the same (in close agreement with the definition of ideal PBR).

Finally, regarding this last comment, it must be pointed out about ideal biomass surface growth rates $<s_X>_{max}$ that if we have identified them with the DiCoFluV concept, it actually exists a technological gap in conceiving and optimizing efficient systems for the collection, concentration and then dilution of incident solar light flux densities inside the culture medium of the PBR (Ogbonna *et al.*, 1999). A considerable optimization work is then needed on this matter in order to develop actual systems with transport efficiency of at least 50%. In the same perspective, it will be also necessary to develop and optimize hybrid PBR with solar collector systems able to separate the infrared radiation (Schlegel *et al.*, 2004) which is not available for photosynthesis (roughly 50% of the total sunlight spectrum) in order to generate electrical power as utility for the reactor. This kind of hybrid system is fully compatible with the DiCoFluV PBR technology and would lead to really high global thermodynamic efficiencies, in the same order of magnitude than those observed for other purely physical conversion of solar radiation (photovoltaic, thermal cycles).

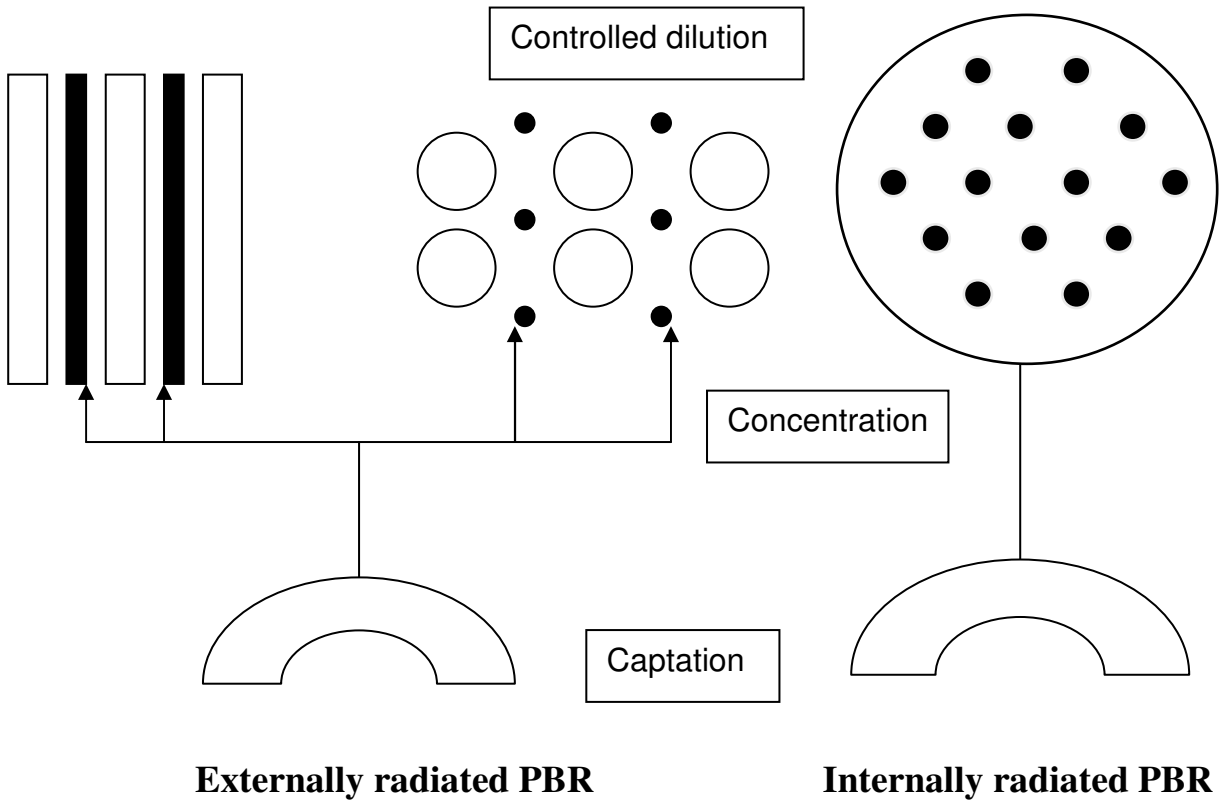

Figure 7: Schematic representation of different kinds of solar DiCoFluV photobioreactors demonstrating that this concept may be declined both with internal position of the lighting structures inside the culture medium (internally radiated, as treated in the article), or with external position of these lighting structures (using transparent walls) in any geometry considered (externally radiated).

## 6- Conclusions and Perspectives

In this article, we have first derived, from simplified engineering equations relying on full knowledge models established in the last twenty years, some basic rules defining the optimal functioning of large size industrial PBR, including surface and volume kinetic productivities as well as energetic performances. As an immediate conclusion, it has then been demonstrated from a purely theoretical point of view that volumetrically-lightened PBR (with artificial or solar light) were the best technologies for many industrial applications if maximum biomass productivities were envisaged. The optimal design of these concepts, mainly regarding the best assembly of internal lighting structures inside the reactor volume has been investigated using the constructal approach. Special attention has been paid on the 2D-cylindrical geometry because of its correspondence with many practical cases, but the method has been further generalized to any other geometry, providing the optimal design of highly efficient PBR, up scalable with constant productivities by geometrical similitude. This successful result enables to extend the field of application of the constructal approach used here for the first time, to our best knowledge, when considering a radiation transfer problem linked to a participative and reactive material phase. For this class of problems, a thermodynamic analysis of the optimization procedure has been proposed and discussed according to the constructal law demonstrating that the better access to flux corresponds to a maximum of the exergetic efficiency for the PBR. Finally, the resulting optimal assemblies obtained for volumetrically-lightened PBR have been compared and identified to pre-established concepts of ideal PBR. Mainly, the DiCoFluV concept has been proved to be the solar ideal PBR as a thermodynamic limit for surface biomass productivities.

Finally, it must be retained that this work opens the way towards the engineering of new and highly efficient large-size PBR on solid, rational and quantitative theoretical bases, but however from simple analytical tools, in the spirit of the constructal approach (Bejan, 2000) and of the Hartmanis principle (Cornet and Dussap, 2009). Alternatively, a rigorous and fully numerical method consisting in studying the domain sensitivity (the geometry) by Monte Carlo calculation of the internally-lightened PBR will be examined in the future from the theoretical tools developed by Roger *et al*. (2005). In any case, the real concepts of PBR which will be inspired by the optimal designs developed in this study will require further technological developments if they intend to reach the ideal performances calculated in this article.

## ACKNOWLEDGEMENTS

The author thanks the Centre National d'Études Spatiales (CNES, France) and the Agence Nationale de la Recherche (ANR, France, BIOSOLIS project) which partly supported this work.

# NOMENCLATURE

| | | |
|---|---|---|
| $a$ | Volumetric absorption coefficient | [m$^{-1}$] |
| $a_{light}$ | Specific illuminated area for any given photobioreactor | [m$^{-1}$] |
| $a_S$ | Specific illuminated area relative lighting structures | [m$^{-1}$] |
| $\mathcal{A}$ | Local volumetric radiant light power density absorbed | [W.m$^{-3}$] |
| $\bar{b}$ | Back-scattered fraction for radiation | [dimensionless] |
| $c_l$ | Generalized one-dimensional abscissa in any system of coordinates | [m] |
| $C_X$ | Biomass concentration | [kg.m$^{-3}$ or g.L$^{-1}$] |
| $C^*$ | Solar constant | [W.m$^{-2}$] |
| $d$ | Internal diameter for a cylindrical annular PBR | [m] |
| $d_i$ | Distance between two lighting structures in any system of coordinates | [m] |
| $d_S$ | Diameter of the lighting structure in any system of coordinates | [m] |
| $D$ | Total diameter for a cylindrical photobioreactor | [m] |
| $Ea$ | Mass absorption coefficient | [m$^2$.kg$^{-1}$] |
| $Es$ | Mass scattering coefficient | [m$^2$.kg$^{-1}$] |
| $f_d$ | Design dark volume fraction of any photobioreactor | [dimensionless] |
| $G$ | Local spherical irradiance | [W.m$^{-2}$] |
| $G_C$ | Value of the irradiance corresponding to the compensation point for photosynthesis | [W.m$^{-2}$] |
| $I$ | Specific radiant intensity | [W.m$^{-2}$] |
| $k_h$ | Define irradiances and light flux densities as a function of the angular distribution of intensities (see eq. B5 for definition) | [dimensionless] |
| $K$ | Half saturation constant for photosynthesis | [W.m$^{-2}$] |
| $L$ | Total length for a rectangular photobioreactor | [m] |
| $m$ | Number of angles defining the generalized frame of reference for the RTE | [dimensionless] |
| $M_X$ | C-molar mass for the biomass | [kg.mol$^{-1}$] |
| $n$ | Degree of collimation for the radiation field | [dimensionless] |
| $p(\mathbf{\Omega},\mathbf{\Omega}')$ | Phase function for scattering | [dimensionless] |
| $\bar{q}_0$ | Mean hemispherical incident light flux density (LFD) in the PAR | [W.m$^{-2}$] |
| $\bar{q}_\perp$ | Mean hemispherical incident light flux density (LFD) in the PAR in the particular case of a solar tracking system (direct solar light only) | [W.m$^{-2}$] |
| $\bar{q}_\cap$ | Mean hemispherical incident light flux density (LFD) in the PAR relative to the normal of a fixed plan of reference (diffuse and direct solar light) | [W.m$^{-2}$] |
| $r$ | Radius in cylindrical geometry | [m] |
| $r_X$ | Local biomass volumetric growth rate (volumetric productivity) | [kg.m$^{-3}$.h$^{-1}$ or t.m$^{-3}$.y$^{-1}$] |
| $s$ | Volumetric scattering coefficient | [m$^{-1}$] |
| $s_X$ | Biomass surface growth rate (surface productivity) | [kg.m$^{-2}$.h$^{-1}$ or t.ha$^{-1}$.y$^{-1}$] |
| $t_S$ | Maximum yearly solar illumination duration | [h] |
| $V$ | Volume of any reactor | [m$^3$ or L] |

**Greek letters**

| | | |
|---|---|---|
| $\alpha$ | Linear scattering modulus | [dimensionless] |
| $\beta, \beta'$ | Zenith (polar) angle | [rad] |
| $\gamma$ | Volume working illuminated fraction inside the photobioreactor | [dimensionless] |
| $\Delta g'^0_X$ | Standard mean Gibbs free energy of formation for biomass | [J.mol$^{-1}$] |
| $\delta$ | Extinction coefficient for the generalized two-flux method | [m$^{-1}$] |
| $\varepsilon$ | Volume fraction of the lighting structures inside the total reactor volume | [dimensionless] |
| $\varepsilon_{max}$ | Maximum volume fraction of the lighting structures inside the total reactor volume | [dimensionless] |
| $\eta_{th}$ | Thermodynamic efficiency (exergetic yield) of the photobioreactor | [dimensionless] |
| $\tilde{\mu}_i$ | Chemical potential of the species $i$ | [J.mol$^{-1}$] |
| $\xi$ | Azimuth angle | [rad] |
| $\rho_M$ | Maximum energetic yield for photon conversion | [dimensionless] |
| $\bar{\tau}$ | Mean (daily integrated) solar transmission of the atmosphere | [dimensionless] |
| $\upsilon_{ij}$ | Stoichiometric coefficient | [dimensionless] |
| $\bar{\psi}$ | Mean mass energetic yield for the Z-scheme of photosynthesis | [kg.J$^{-1}$] |
| $\Omega$ | Solid angle | [rad] |
| $\boldsymbol{\Omega}$ | Unit directional vector | [dimensionless] |

**Subscripts**

| | |
|---|---|
| max | Relative to maximum values (mainly for biomass productivities) |
| opt | Relative to optimal values (mainly for the characteristic distances $d_i$ and $d_S$) |
| $X$ | Relative to biomass dry weight |
| $\lambda$ | Relative to a spectral quantity for the wavelength $\lambda$ |
| 0 | Relative to the input surface of a rectangular photobioreactor |
| $\perp$ | Relative to a mean daily averaged solar quantity with moving orientation of the normal (direct collection of sun with tracking) |
| $\cap$ | Relative to a mean daily averaged solar quantity respectively to a fixed normal of reference |

**Superscripts**

| | |
|---|---|
| + | Relative to the positive hemisphere in defining the radiation field (positive cos $\beta$) |
| - | Relative to the negative hemisphere in defining the radiation field (negative cos $\beta$) |

**Other**

$$\bar{X} = \frac{1}{\Delta t} \int_{\Delta t} X \, dt \qquad \text{Time averaging}$$

$$<X> = \frac{1}{V} \iiint_V X \, dV \qquad \text{Spatial averaging}$$

**Abbreviations**

| | |
|---|---|
| AM | Air mass |
| DiCoFluV | Dilution contrôlée du flux en volume |
| DOF | Degree of freedom |
| HPV | Haute productivité volumique |

LFD        Light flux density

PAR        Photosynthetically active radiation

PBR        Photobioreactor

RTE        Radiative transfer equation

Post print published in Chemical Engineering Science, 65: 985-998, 2010

## APPENDIX A

Calculation of Ideal Hemispherical Incident Sunlight Fluxes in the PAR for Ideal Performances of Solar PBR

The calculation of ideal kinetic performances of solar PBR requires defining first ideal daily sunlight conditions in the PAR and ideal yearly illumination conditions. In the following, these quantities are assessed from the assumptions of a PBR located at the equator and at the sea level, with an ideal atmosphere.

The mean direct hemispherical radiant light flux density $\overline{q}_\perp$ (with a normal coinciding to the direction of the sun) on a typical day can be first obtained from the value of the solar constant outside atmosphere $C^* = 1367$ W/m$^2$ (WMO normalization). Using then a relation for atmosphere attenuation at a given air mass (a function of the zenithal angle) as proposed by Bernard (2004) and averaging on a diurnal time course of the sun enables to calculate the mean transmission as:

$$\overline{\tau} \cong \frac{1}{2\pi}\left[ G_{10}^{30}\left(0.11 \,\Big|_{1/2,1/2,0}^{1}\right) + G_{10}^{30}\left(0.0023 \,\Big|_{1/2,1/2,0}^{1}\right) \right] \cong 0.57$$

where $G_{pq}^{mn}\left(x \,\Big|_{b(1),...,b(q)}^{a(1),...,a(p)}\right)$ is the G Meijer function. Finally, retaining that the fraction of the total solar radiation at the earth surface in the PAR [400-700 nm] is 43.3% (for AM = 1.5G, IEC 904-3 reference) leads to the ideal value of the mean direct hemispherical light flux density for PBR applications:

$$\overline{q}_\perp = \overline{\tau} \times 0.433 \times C^* \cong 340 \text{ W/m}^2$$

This value corresponds to the maximum daily-averaged hemispherical light flux density (PAR) available at the earth surface if using a tracking system of the sun (only operating with direct radiation). It will be used along with this article when referring to the DiCoFluV concept.

At the opposite, if no tracking system exists, i.e. for a classical PBR with passive solar illumination through transparent walls, it becomes necessary to cosine average again the incoming radiation on a horizontal reference surface (in the ideal case of equator) to obtain the mean global hemispherical light flux density $\bar{q}_\cap$. In this case nevertheless, the PBR can use both the direct and diffuse radiation but this later can be obtained in ideal conditions using the AM = 1.5G (IEC 904-3 reference) reference. Thus, the maximum daily-averaged hemispherical light flux density (PAR) available on a reference surface is roughly given by:

$$\bar{q}_\cap = \frac{2}{\pi} \times 1.17 \times \bar{q}_\perp \cong 250 \text{ W/m}^2$$

It will be used in this article when referring to ideal classical PBR directly illuminated by the sun.

Additionally, for these two cases, in order to calculate yearly productivities of PBR, the ideal sunlight yearly duration will be simply taken as $t_s \cong 12 \times 365 = 4380$ h. This value is very close to the 4300 h actually measured in some oriental countries of Sahara (Bernard, 2004).

Evidently, if the PBR operates at any other location, these values must be corrected by accurate gnomonic calculations, using data banks or roughly correcting by the latitude and the irradiation factor (Bernard, 2004). In all cases, the values of fluxes and sunlight duration will be lower so leading to decrease the ideal biomass productivities calculated in this article.

## APPENDIX B

The Simplified Quasi One-dimensional Approach to Solve the Radiative Transfer Equation in Rectangular and Curvilinear Systems of Coordinates for the Calculation of the Optimal Distance $(d_i)_{opt}$ - First Construct

In this appendix, we present a general method in solving analytically the one-dimensional form of the radiative transfer equation (RTE) in any system of coordinates. For sake of generality, we adopt here a general spectral notation (using the indices $\lambda$), but in any case, the problem is the same if using the spectrally-averaged form of the RTE (requiring to have mean radiative properties for the material phase in the PAR as it is considered in the core of this paper, i.e. mean volumetric absorption $a$ and scattering $s$ coefficients).

| Cartesian system | Cylindrical system | Spherical system |
|---|---|---|
| 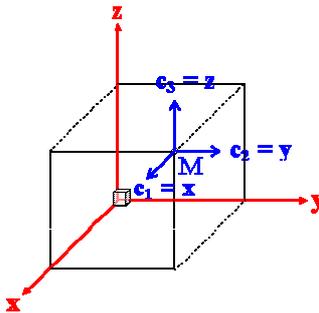 | 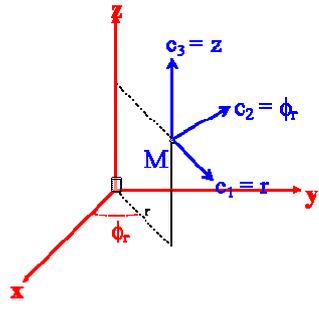 | 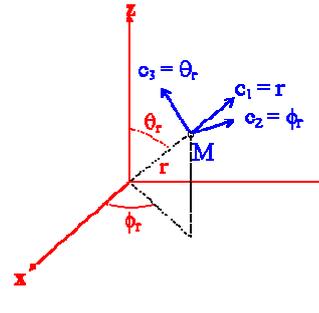 |
| 3 abcissas ($x$, $y$, $z$) | 2 abcissas ($r$, $z$) + 1 angle ($\phi_r$) | 1 abscissa ($r$) + 2 angles ($\phi_r$, $\theta_r$) |
| **$m = 0$** | **$m = 1$** | **$m = 2$** |

<u>Scheme B1</u>: definitions of the fixed frame of reference for the three considered systems of coordinates. The number of abscissas and angles (defining the parameter $m$) are also presented in each case.

This problem requires first to examine the status of the operator for the quasi steady-sate three dimensional form of the RTE given by (Cassano *et al*., 1995; Siegel and Howell, 2002):

$$\nabla \cdot (\mathbf{\Omega} I_\lambda) = -(a_\lambda + s_\lambda) I_\lambda + \frac{s_\lambda}{4\pi} \int_{\beta'=0}^{\pi} \int_{\xi=0}^{2\pi} I_\lambda p_\lambda (\mathbf{\Omega}, \mathbf{\Omega}') \sin\beta' d\beta' d\xi' \quad \text{(B1)}$$

$\nabla \cdot (\mathbf{\Omega} I_\lambda)$ is in fact a five Euclidean dimensional operator relying first on a fixed frame of reference with three parameters depending on the geometry of the problem (1 to 3 abscissas and $m = 0$ to 2 angles as depicted on scheme B1), and on a moving frame of reference (defining the unit vector $\mathbf{\Omega}$) with two parameters independent of the considered geometry as depicted on scheme B2.

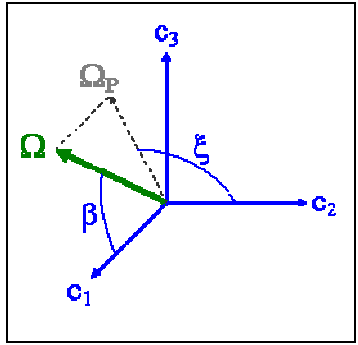

Scheme B2: definition of the zenith $\beta$ and azimuthal $\xi$ angles for any considered geometry, then characterizing the unit vector $\mathbf{\Omega}$ respectively to the reference abscissa $c_1$.

With this choice, the unit vector $\mathbf{\Omega}$ is characterized by two angles, the zenith angle $\beta$ and the azimuthal angle $\xi$, defined from the coordinate $c_1$, considered as a key coordinate in any investigated geometry. This enables to always define the one-dimensional operator of the RTE in the same manner (as proposed by Van Noort et al., (2002) for a more complicated two-dimensional problem) for any geometry, independently of the azimuth, and leading to:

$$\nabla^\circ \cdot (\mathbf{\Omega} I_\lambda) = \frac{dI_\lambda}{dc_1} \cos \beta \quad \text{(B2)}$$

Eq. (B1) can then be analytically solved only if it is separated in a pair of equations on each hemisphere, as it is the case in the well-known rectangular two-flux approach (Siegel and Howell, 2002). This method has been historically proposed in Cartesian coordinates by Schuster (1905) for a diffuse radiation and latter extended to a collimated radiation (Hottel and Sarofim, 1967). In this article, we generalized this approach, not only to any system of coordinates as already explained, but also to a more general case for the angular dependence of the intensities $I_\lambda$. We postulate in fact a general angular dependence for each hemisphere in the form:

$$I_\lambda^\pm(\beta) = I_\lambda^\pm \left| \cos^n \beta \right| \quad \text{(B3)}$$

in which $n$ is called the degree of collimation (leading to the isotropic case with $n = 0$ and to the collimated case with $n = \infty$). Thus, using this definition in the so-called pair of equations derived from one-dimensional eqs. (B1) and (B2) and integrating over each hemisphere (positive and negative values of $\cos \beta$) with the help of summation rules for the $n^{th}$ moment equations obtained, leads to a first independent relation for the intensities in the form:

$$\frac{d\left(I_\lambda^+ + I_\lambda^-\right)}{d c_1} = -k_h (a_\lambda + 2\bar{b}_\lambda s_\lambda)\left(I_\lambda^+ - I_\lambda^-\right) \quad \text{(B4)}$$

in which the collimation factor $k_h$ and the back scattered fraction $\bar{b}_\lambda$ are defined from the phase function for scattering $p(\beta, \beta')$ by:

$$\begin{aligned} k_h &= \frac{n+2}{n+1} \\ \bar{b}_\lambda &= \frac{1}{2} \int_{\frac{\pi}{2}}^{\pi} p_\lambda(\beta, \beta') \sin \beta \, d\beta \end{aligned} \quad \text{(B5)}$$

Additionally, a second independent equation is required, mainly in curvilinear systems of coordinates, in order to be sure that the obtained solutions verify the energy balance on the photonic phase. From our previous notations and assumptions, this balance can be written for a non-emitting medium (Chandrasekhar, 1960; Siegel and Howell, 2002) as:

$$\frac{d(I_\lambda^+ - I_\lambda^-)}{dc_1} + \frac{m}{c_1}(I_\lambda^+ - I_\lambda^-) = -k_h a_\lambda (I_\lambda^+ + I_\lambda^-) \quad (B6)$$

in which the parameter *m*, depending on the system of coordinates, has been introduced on scheme A1. Finally, differentiating eq. (B4) and combining with eq. (B6) leads to the governing master equation for the one-dimensional two-flux approximation RTE in any geometry:

$$\frac{d^2(I_\lambda^+ + I_\lambda^-)}{dc_1^2} + \frac{m}{c_1}\frac{d(I_\lambda^+ + I_\lambda^-)}{dc_1} - k_h^2 a_\lambda (a_\lambda + 2\bar{b}_\lambda s_\lambda)(I_\lambda^+ + I_\lambda^-) = 0 \quad (B7)$$

or in alternative form, defining the spectral irradiance $G_\lambda = \iint_{4\pi} I_\lambda \, d\Omega \equiv k_h \pi^{k_h - 1}(I_\lambda^+ + I_\lambda^-)$:

$$\frac{d^2 G_\lambda}{dc_1^2} + \frac{m}{c_1}\frac{d G_\lambda}{dc_1} - k_h^2 a_\lambda (a_\lambda + 2\bar{b}_\lambda s_\lambda) G_\lambda = 0$$

Evidently, eq. (B7) requires appropriate boundary conditions to be solved as examined for example by Takache *et al.* (2009), but this simple mathematical form can lead to analytical solutions in numerous practical cases. This is of crucial importance in the framework of finding the optimal distance $(d_i)_{opt}$ of the first construct, each time that the quasi-one-dimensional assumption can be done for the smallest scale of the optimization procedure. In other situations indeed, eq. (B1) must be numerically solved using appropriate methods for photon transport equations such as characteristic-scheme finite element methods (Cornet *et al.*, 1994; Van Noort *et al.*, 2002) or Monte Carlo methods (Csogör *et al.*, 2001; Siegel and Howell, 2002), complicating dramatically the calculation of $(d_i)_{opt}$.

## APPENDIX C

The General Geometrical Surface-to-Volume Optimization Problem in Rectangular (1D), Cylindrical (2D) or Spherical (3D) Systems for the calculation of $(d_S)_{opt}$ - Second Construct

The aim of this appendix is to generalize the geometrical surface-to-volume optimization problem in finding the optimal lighting structure diameter $(d_S)_{opt}$, as the last step of the constructal approach, in any geometry. As demonstrated in the core of this article, this optimization problem requires maximizing the mean volumetric biomass growth rate $<r_X>$ (see eq. 1), once the optimization of the distance $(d_i)_{opt}$ at the smaller scale has been done. This last problem (the first construct) just requires a theoretical numerical or analytical tool to analyze the radiant light transport phenomena in a given geometry and coordinates system, as provided for example by the approach depicted in appendix B and in the core of the article. This leads to the establishment of drawings like Fig. 4 giving the resulting optimal value of $(d_i)_{opt}$ versus the biomass concentration $C_X$ and the incident LFD $\bar{q}_0$. Here, this step is assumed already performed in any geometry and we focus our analysis on the final level of optimization, at the scale of the PBR itself.

It has been already discussed that it was possible to use, as a very good approximation, the reliable general formula recently published (Cornet and Dussap, 2009) and giving the maximum ($\gamma = 1$) volumetric growth rate of a PBR, illuminated with an incident LFD $\bar{q}_0$, in any geometry. Redefining thus the specific illuminated area of the reactor $a_{light}$ from the specific structures surface $a_S$ and their volume fraction $\varepsilon$ relative to the total volume of the reactor, leads then easily to the general relationship for maximum kinetic performances, including the special case of (2D-problem) cylindrical geometry (Cornet and Dussap, 2009):

$$<r_X> \equiv \rho_M \overline{\psi} \frac{2\alpha}{1+\alpha} a_S \varepsilon \frac{K}{k_h} \ln\left[1+\frac{k_h \overline{q_0}}{K}\right] \quad (C1)$$

in which the energetic yields $\rho_M$ and $\overline{\psi}$ are known by a predictive approach, $k_h$ is defined in appendix B (eq. B5), the dark engineering fraction $f_d$ is taken to zero, and the two previous parameters are given from the general geometrical definition of *m* (see appendix B, scheme B1) by:

$$a_S = \frac{2(m+1)}{d_S}$$

$$\varepsilon = \frac{\varepsilon_{max}}{\left(1+\frac{d_i}{d_S}\right)^{m+1}} \quad (C2)$$

Equations (C1-C2) are in fact general relationships generating the three optimization problems corresponding to the three considered geometries (Cartesian with *m* = 0, cylindrical with *m* = 1 and spherical with *m* = 2) and enabling to maximize the lighting surface in a given volume for each situation. The Cartesian problem first corresponds to the case of multi-layered rectangular PBR illuminated by alternating thin and flat lateral light diffusing panels with rectangular culture chambers of thickness $d_i$ (see Fig. 7 for an example of such externally radiated PBR). This is a one-dimensional problem because the height and the width of the panel have no effect on the design (they are only purely technical factors) and the optimization concerns only the arrangement of many panels in a given direction (one axis *x*). The cylindrical problem has been extensively studied in this article; it corresponds indeed to a two-dimensional problem because the optimization process supposes finding the best array of small lateral light diffusing circles inside a given surface (the cross section of the PBR, cylindrical or not, then requiring two axes *x* and *y* to be constructed). Finally, the spherical geometry for lighting structures is a three-dimensional problem because it requires to find the best assembly of small radiating spheres in a given volume (the entire PBR, involving three axis *x*, *y*, *z* to be defined). It could appear here as a scholar, purely theoretical case study, but actually, such system can be imagined, using luminescent particles in the volume of the reactor. For example, the author has developed a small completely stirred photoreactor with two (emulsion-like) liquid-liquid phases operating in continuous functioning, with a chemo-luminescent reaction taking place in the organic phase (using mainly 5,12-

bis(phenyl-ethynyl)-naphtacene and bis(2,4,6-trichlorophenyl)oxalate) and providing an incident LFD $\bar{q}_0$ of roughly 5 W/m² to the aqueous phase with a very high specific area $a_S$ (data not shown).

In all cases, the maximum volume fraction $\varepsilon_{max}$ appearing in eq. (C2) corresponds to the situation where the lighting structures touch each other in any geometry, and then has well-known values given by:

$$\text{- Cartesian coordinates, 1D - problem,} \quad m = 0, \quad \varepsilon_{max} = 1$$

$$\text{- Cylindrical coordinates, 2D - problem,} \quad m = 1, \quad \varepsilon_{max} = \frac{\pi}{2\sqrt{3}} \quad (C3)$$

$$\text{- Spherical coordinates, 3D - problem,} \quad m = 2, \quad \varepsilon_{max} = \frac{\pi\sqrt{2}}{6}$$

The general optimization problem can then be solved using eqs. (C1-C3) and applying the criterion (with a distance $(d_i)_{opt}$ optimized at the level of the first construct):

$$<r_X>_{max} \implies \left(\frac{\partial <r_X>}{\partial d_S}\right)\bigg|_{C_X, \bar{q}_0, (d_i)_{opt}} = 0 \quad (C4)$$

leading very easily to the final values for any geometry:

$$(d_S)_{opt} = m\,(d_i)_{opt} \quad (C5)$$

and to the respective optimal lighting structures volume fractions from eq. (C2):

$$\text{- Cartesian coordinates, 1D - problem,} \quad m = 0, \quad \varepsilon_{opt} \to 0$$

$$\text{- Cylindrical coordinates, 2D - problem,} \quad m = 1, \quad \varepsilon_{opt} = \frac{\pi}{8\sqrt{3}} \quad (C6)$$

$$\text{- Spherical coordinates, 3D - problem,} \quad m = 2, \quad \varepsilon_{opt} = \frac{4\pi\sqrt{2}}{81}$$

These important (and simple) results enable finally to use eqs. (C1-C2) and eqs. (C5-C6) to calculate and plot the dimensionless volumetric growth rates (relative to the maximum value always given by the rectangular case with $(d_S/d_i) \to 0$ or $\varepsilon \to 0$) versus the dimensionless $(d_S/d_i)$ abscissa, as it has been done on Fig. 6, confirming the optimal values respectively given by eq. (C5).